# Climate Science: Is it currently designed to answer questions?[1]

Richard S. Lindzen
Program in Atmospheres, Oceans and Climate
Massachusetts Institute of Technology

November 29, 2008
Revised: September 21, 2012

## Abstract

For a variety of inter-related cultural, organizational, and political reasons, progress in climate science and the actual solution of scientific problems in this field have moved at a much slower rate than would normally be possible. Not all these factors are unique to climate science, but the heavy influence of politics has served to amplify the role of the other factors. By cultural factors, I primarily refer to the change in the scientific paradigm from a dialectic opposition between theory and observation to an emphasis on simulation and observational programs. The latter serves to almost eliminate the dialectical focus of the former. Whereas the former had the potential for convergence, the latter is much less effective. The institutional factor has many components. One is the inordinate growth of administration in universities and the consequent increase in importance of grant overhead. This leads to an emphasis on large programs that never end. Another is the hierarchical nature of formal scientific organizations whereby a small executive council can speak on behalf of thousands of scientists as well as govern the distribution of 'carrots and sticks' whereby reputations are made and broken. The above factors are all amplified by the need for government funding. When an issue becomes a vital part of a political agenda, as is the case with climate, then the politically desired position becomes a goal rather than a consequence of scientific research. This paper will deal with the origin of the cultural changes and with specific examples of the operation and interaction of these factors. In particular, we will show how political bodies act to control scientific institutions, how scientists adjust both data and even theory to accommodate politically correct positions, and how opposition to these positions is disposed of.

---

# 1. Introduction.

Although the focus of this paper is on climate science, some of the problems pertain to science more generally. Science has traditionally been held to involve the creative opposition of theory and observation wherein each tests the other in such a manner as to converge on a better understanding of the natural world. Success was rewarded by recognition, though the degree of recognition was weighted according to both the practical consequences of the success and the philosophical and aesthetic power of the success. As science undertook more ambitious problems, and the cost and scale of operations increased, the need for funds undoubtedly shifted emphasis to practical relevance though numerous examples from the past assured a strong base level of confidence in the utility of science. Moreover, the many success stories established 'science' as a source of authority and integrity. Thus, almost all modern movements claimed scientific foundations for their aims. Early on, this fostered a profound misuse of science, since science is primarily a successful mode of inquiry rather than a source of authority.

Until the post World War II period, little in the way of structure existed for the formal support of science by government (at least in the US which is where my own observations are most relevant). In the aftermath of the Second World War, the major contributions of science to the war effort (radar, the A-bomb), to health (penicillin), etc. were evident. Vannevar Bush (in his report, *Science: The Endless Frontier,* 1945) noted the many practical roles that validated the importance of science to the nation, and argued that the government need only adequately support basic science in order for further benefits to emerge. The scientific community felt this paradigm to be an entirely appropriate response by a grateful nation. The next 20 years witnessed truly impressive scientific productivity which firmly established the United States as the creative center of the scientific world. The Bush paradigm seemed amply justified. (This period and its follow-up are also discussed by Miller, 2007, with special but not total emphasis on the NIH (National Institutes of Health).) However, something changed in the late 60's. In a variety of fields it has been suggested that the rate of new discoveries and achievements slowed appreciably (despite increasing publications)[2], and it is being suggested that either the Bush paradigm ceased to be valid or that it may never have been valid in the first place. I believe that the former is correct. What then happened in the 1960's to produce this change?

---

[2] At some level, this is obvious. Theoretical physics is still dealing with the standard model though there is an active search for something better. Molecular biology is still working off of the discovery of DNA. Many of the basic laws of physics resulted from individual efforts in the 17th-19th Centuries. The profound advances in technology should not disguise the fact that the bulk of the underlying science is more than 40 years old. This is certainly the case in the atmospheric and oceanic sciences. That said, it should not be forgotten that sometimes progress slows because the problem is difficult. Sometimes, it slows because the existing results are simply correct as is the case with DNA. Structural problems are not always the only factor involved.

### *Climate Science: Is it currently designed to answer questions?*

It is my impression that by the end of the 60's scientists, themselves, came to feel that the real basis for support was not gratitude (and the associated trust that support would bring further benefit) but fear: fear of the Soviet Union, fear of cancer, etc. Many will conclude that this was merely an awakening of a naive scientific community to reality, and they may well be right. However, between the perceptions of gratitude and fear as the basis for support lies a world of difference in incentive structure. If one thinks the basis is gratitude, then one obviously will respond by contributions that will elicit more gratitude. The perpetuation of fear, on the other hand, militates against solving problems. This change in perception proceeded largely without comment. However, the end of the cold war, by eliminating a large part of the fear-base forced a reassessment of the situation. Most thinking has been devoted to the emphasis of other sources of fear: competitiveness, health, resource depletion and the environment.

What may have caused this change in perception is unclear, because so many separate but potentially relevant things occurred almost simultaneously. The space race reinstituted the model of large scale focused efforts such as the moon landing program. For another, the 60's saw the first major postwar funding cuts for science in the US. The budgetary pressures of the Vietnam War may have demanded savings someplace, but the fact that science was regarded as, to some extent, dispensable, came as a shock to many scientists. So did the massive increase in management structures and bureaucracy which took control of science out of the hands of working scientists. All of this may be related to the demographic pressures resulting from the baby boomers entering the workforce and the post-sputnik emphasis on science. Sorting this out goes well beyond my present aim which is merely to consider the consequences of fear as a perceived basis of support.

Fear has several advantages over gratitude. Gratitude is intrinsically limited, if only by the finite creative capacity of the scientific community. Moreover, as pointed out by a colleague at MIT, appealing to people's gratitude and trust is usually less effective than pulling a gun. In other words, fear can motivate greater generosity. Sputnik provided a notable example in this regard; though it did not immediately alter the perceptions of most scientists, it did lead to a great increase in the number of scientists, which contributed to the previously mentioned demographic pressure. Science since the sixties has been characterized by the large programs that this generosity encourages. Moreover, the fact that fear provides little incentive for scientists to do anything more than perpetuate problems, significantly reduces the dependence of the scientific enterprise on unique skills and talents. The combination of increased scale and diminished emphasis on unique talent is, from a certain point of view, a devastating combination which greatly increases the potential for the political direction of science, and the creation of dependent constituencies. With these new constituencies, such obvious controls as peer review and detailed accountability, begin to fail and even serve to perpetuate the defects of the system. Miller (2007) specifically addresses how the system especially favors dogmatism and conformity.

The creation of the government bureaucracy, and the increasing body of regulations accompanying government funding, called, in turn, for a massive increase in the administrative staff at universities and research centers. The support for this staff comes from the overhead on

*Climate Science: Is it currently designed to answer questions?*

government grants, and, in turn, produces an active pressure for the solicitation of more and larger grants[3].

One result of the above appears to have been the deemphasis of theory because of its intrinsic difficulty and small scale, the encouragement of simulation instead (with its call for large capital investment in computation), and the encouragement of large programs unconstrained by specific goals[4]. In brief, we have the new paradigm where simulation and programs have replaced theory and observation, where government largely determines the nature of scientific activity, and where the primary role of professional societies is the lobbying of the government for special advantage.

This new paradigm for science and its dependence on fear based support may not constitute corruption *per se*, but it does serve to make the system particularly vulnerable to corruption. Much of the remainder of this paper will illustrate the exploitation of this vulnerability in the area of climate research. The situation is particularly acute for a small weak field like climatology. As a field, it has traditionally been a subfield within such disciplines as meteorology, oceanography, geography, geochemistry, etc. These fields, themselves are small and immature. At the same time, these fields can be trivially associated with natural disasters. Finally, climate science has been targeted by a major political movement, environmentalism, as the focus of their efforts, wherein the natural disasters of the earth system, have come to be identified with man's activities – engendering fear as well as an agenda for societal reform and control. The remainder of this paper will briefly describe how this has been playing out with respect to the climate issue.

---

[3] It is sometimes thought that government involvement automatically implies large bureaucracies, and lengthy regulations. This was not exactly the case in the 20 years following the second world war. Much of the support in the physical sciences came from the armed forces for which science support remained a relatively negligible portion of their budgets. For example, meteorology at MIT was supported by the Air Force. Group grants were made for five year periods and renewed on the basis of a site visit. When the National Science Foundation was created, it functioned with a small permanent staff supplemented by 'rotators' who served on leave from universities for a few years. Unfortunately, during the Vietnam War, the US Senate banned the military from supporting non-military research (Mansfield Amendment). This shifted support to agencies whose sole function was to support science. That said, today all agencies supporting science have large 'supporting' bureaucracies.

[4] In fairness, such programs should be distinguished from team efforts which are sometimes appropriate and successful: classification of groups in mathematics, human genome project, etc.

*Climate Science: Is it currently designed to answer questions?*

## 2. Conscious Efforts to Politicize Climate Science

The above described changes in scientific culture were both the cause and effect of the growth of 'big science,' and the concomitant rise in importance of large organizations. However, all such organizations, whether professional societies, research laboratories, advisory bodies (such as the national academies), government departments and agencies (including NASA, NOAA, EPA, NSF, etc.), and even universities are hierarchical structures where positions and policies are determined by small executive councils or even single individuals. This greatly facilitates any conscious effort to politicize science via influence in such bodies where a handful of individuals (often not even scientists) speak on behalf of organizations that include thousands of scientists, and even enforce specific scientific positions and agendas. The temptation to politicize science is overwhelming and longstanding. Public trust in science has always been high, and political organizations have long sought to improve their own credibility by associating their goals with 'science' – even if this involves misrepresenting the science.

Professional societies represent a somewhat special case. Originally created to provide a means for communication within professions – organizing meetings and publishing journals – they also provided, in some instances, professional certification, and public outreach. The central offices of such societies were scattered throughout the US, and rarely located in Washington. Increasingly, however, such societies require impressive presences in Washington where they engage in interactions with the federal government. Of course, the nominal interaction involves lobbying for special advantage, but increasingly, the interaction consists in issuing policy and scientific statements on behalf of the society. Such statements, however, hardly represent independent representation of membership positions. For example, the primary spokesman for the American Meteorological Society in Washington is Anthony Socci who is neither an elected official of the AMS nor a contributor to climate science. Rather, he is a former staffer for Al Gore.

Returning to the matter of scientific organizations, we find a variety of patterns of influence. The most obvious to recognize (though frequently kept from public view), consists in prominent individuals within the environmental movement simultaneously holding and using influential positions within the scientific organization. Thus, John Firor long served as administrative director of the National Center for Atmospheric Research in Boulder, Colorado. This position was purely administrative, and Firor did not claim any scientific credentials in the atmospheric sciences at the time I was on the staff of NCAR. However, I noticed that beginning in the 1980's, Firor was frequently speaking on the dangers of global warming as an expert from NCAR. When Firor died last November, his obituary noted that he had also been Board Chairman at Environmental Defense– a major environmental advocacy group – from 1975-1980[5]. The UK Meteorological Office also has a board, and its chairman, Robert Napier, was

---

[5] A personal memoir from Al Grable sent to Sherwood Idso in 1993 is interesting in this regard. Grable served as a Department of Agriculture observer to the National Research Council's National Climate Board. Such observers are generally posted by agencies to boards

*Climate Science: Is it currently designed to answer questions?*

previously the Chief Executive for World Wildlife Fund - UK.  Bill Hare, a lawyer and Campaign Director for Greenpeace, frequently speaks as a 'scientist' representing the Potsdam Institute, Germany's main global warming research center.  John Holdren, who currently directs the Woods Hole Research Center (an environmental advocacy center not to be confused with the far better known Woods Hole Oceanographic Institution, a research center), is also a professor in Harvard's Kennedy School of Government, and has served as president of the American Association for the Advancement of Science among numerous other positions including serving on the board of the MacArthur Foundation from 1991 until 2005.  He was also a Clinton-Gore Administration spokesman on global warming.  The making of academic appointments to global warming alarmists is hardly a unique occurrence.  The case of Michael Oppenheimer is noteworthy in this regard.  With few contributions to climate science (his postdoctoral research was in astro-chemistry), and none to the physics of climate, Oppenheimer became the Barbara Streisand Scientist at Environmental Defense[6].  He was subsequently appointed to a professorship at Princeton University, and is now, regularly, referred to as a prominent climate scientist by Oprah (a popular television hostess), NPR (National Public Radio), etc.  To be sure, Oppenheimer did coauthor an early absurdly alarmist volume (Oppenheimer and Robert Boyle, 1990: *Dead Heat, The Race Against the Greenhouse Effect*), and he has served as a lead author with the IPCC (Intergovernmental Panel on Climate Change)[7].

---

that they are funding.  In any event, Grable describes a motion presented at a Board meeting in 1980 by Walter Orr Roberts, the director of the National Center for Atmospheric Research, and by Joseph Smagorinsky, director of NOAA's Geophysical Fluid Dynamics Laboratory at Princeton, to censure Sherwood Idso for criticizing climate models with high sensitivities due to water vapor feedbacks (in the models), because of their inadequate handling of cooling due to surface evaporation.  A member of that board, Sylvan Wittwer, noted that it was not the role of such boards to censure specific scientific positions since the appropriate procedure would be to let science decide in the fullness of time, and the matter was dropped.  In point of fact, there is evidence that models do significantly understate the increase of evaporative cooling with temperature (Held and Soden, 2006).  Moreover, this memoir makes clear that the water vapor feedback was considered central to the whole global warming issue from the very beginning.

[6] It should be acknowledged that Oppenheimer has quite a few papers with climate in the title – especially in the last two years.  However, these are largely papers concerned with policy and advocacy, assuming significant warming.  Such articles probably constitute the bulk of articles on climate.  It is probably also fair to say that such articles contribute little if anything to understanding the phenomenon.

[7] Certain names and organizations come up repeatedly in this paper.  This is hardly an accident.  In 1989, following the public debut of the issue in the US in Tim Wirth's and Al Gore's famous Senate hearing featuring Jim Hansen associating the warm summer of 1988 with global warming, the Climate Action Network was created.  This organization of over 280 ENGO's has been at the center of the climate debates since then.  The Climate Action Network, is an umbrella NGO that coordinates the advocacy efforts of its members, particularly in relation to the UN negotiations.  Organized around seven regional nodes in North and Latin America, Western and Eastern Europe, South and Southeast Asia, and Africa, CAN represents the majority of environmental groups advocating on climate change, and it has embodied the voice of the

## *Climate Science: Is it currently designed to answer questions?*

One could go on at some length with such examples, but a more common form of infiltration consists in simply getting a couple of seats on the council of an organization (or on the advisory panels of government agencies). This is sufficient to veto any statements or decisions that they are opposed to. Eventually, this enables the production of statements supporting their position – if only as a *quid pro quo* for permitting other business to get done. Sometimes, as in the production of the 1993 report of the NAS, *Policy Implications of Global Warming*, the environmental activists, having largely gotten their way in the preparation of the report where they were strongly represented as 'stake holders,' decided, nonetheless, to issue a minority statement suggesting that the NAS report had not gone 'far enough.' The influence of the environmental movement has effectively made support for global warming, not only a core element of political correctness, but also a requirement for the numerous prizes and awards given to scientists. That said, when it comes to professional societies, there is often no need at all for overt infiltration since issues like global warming have become a part of both political correctness and (in the US) partisan politics, and there will usually be council members who are committed in this manner.

The situation with America's National Academy of Science is somewhat more complicated. The Academy is divided into many disciplinary sections whose primary task is the nomination of candidates for membership in the Academy[8]. Typically, support by more than 85% of the membership of any section is needed for nomination. However, once a candidate is elected, the candidate is free to affiliate with any section. The vetting procedure is generally rigorous, but for over 20 years, there was a Temporary Nominating Group for the Global Environment to provide a back door for the election of candidates who were environmental activists, bypassing the environmental community in the climate negotiations since it was established.

The founding of the Climate Action Network can be traced back to the early involvement of scientists from the research ENGO community. These individuals, including Michael Oppenheimer from Environmental Defense, Gordon Goodman of the Stockholm Environmental Institute (formerly the Beijer Institute), and George Woodwell of the Woods Hole Research Center were instrumental in organizing the scientific workshops in Villach and Bellagio on 'Developing Policy Responses to Climate Change' in 1987 as well as the Toronto Conference on the Changing Atmosphere in June 1988. It should be noted that the current director of the Woods Hole Research Center is John Holdren. In 1989, several months after the Toronto Conference, the emerging group of climate scientists and activists from the US, Europe, and developing countries were brought together at a meeting in Germany, with funding from Environmental Defense and the German Marshall Fund. The German Marshall Fund is still funding NGO activity in Europe:
http://www.gmfus.org/event/detail.cfm?id=453&parent_type=E (Pulver, 2004).

[8] The reports attributed to the National Academy are not, to any major extent, the work of Academy Members. Rather, they are the product of the National Research Council, which consists in a staff of over 1000 who are paid largely by the organizations soliciting the reports. The committees that prepare the reports are mostly scientists who are not Academy Members, and who serve without pay.

*Climate Science: Is it currently designed to answer questions?*

conventional vetting procedure. Members, so elected, proceeded to join existing sections where they hold a veto power over the election of any scientists unsympathetic to their position. Moreover, they are almost immediately appointed to positions on the executive council, and other influential bodies within the Academy. One of the members elected via the Temporary Nominating Group, Ralph Cicerone, is now president of the National Academy. Prior to that, he was on the nominating committee for the presidency. It should be added that there is generally only a single candidate for president. Others elected to the NAS via this route include Paul Ehrlich, James Hansen, Steven Schneider, John Holdren and Susan Solomon.

It is, of course, possible to corrupt science without specifically corrupting institutions. For example, the environmental movement often cloaks its propaganda in scientific garb without the aid of any existing scientific body. One technique is simply to give a name to an environmental advocacy group that will suggest to the public, that the group is a scientific rather than an environmental group. Two obvious examples are the Union of Concerned Scientists and the Woods Hole Research Center[9,10]. The former conducted an intensive advertising campaign about ten years ago in which they urged people to look to them for authoritative information on global warming. This campaign did not get very far – if only because the Union of Concerned Scientists had little or no scientific expertise in climate. A possibly more effective attempt along these lines occurred in the wake of Michael Crichton's best selling adventure, *Climate of Fear*, which pointed out the questionable nature of the global warming issue, as well as the dangers to society arising from the exploitation of this issue. Environmental Media Services (a project of Fenton Communications, a large public relations firm serving left wing and environmental causes; they are responsible for the alar scare as well as Cindy Sheehan's anti-war campaign.) created a website, realclimate.org, as an 'authoritative' source for the 'truth' about climate. This time, real scientists who were also environmental activists, were recruited to organize this web site and 'discredit' any science or scientist that questioned catastrophic anthropogenic global warming. The web site serves primarily as a support group for believers in catastrophe, constantly reassuring them that there is no reason to reduce their worrying. Of course, even the above represent potentially unnecessary complexity compared to the longstanding technique of simply publicly claiming that all scientists agree with whatever catastrophe is being promoted. *Newsweek* already made such a claim in 1988. Such a claim serves at least two purposes. First,

---

[9] One might reasonably add the Pew Charitable Trust to this list. Although they advertise themselves as a neutral body, they have merged with the National Environmental Trust, whose director, Philip Clapp, became deputy managing director of the combined body. Clapp (the head of the legislative practice of a large Washington law firm, and a consultant on mergers and acquisitions to investment banking firms), according to his recent obituary, was 'an early and vocal advocate on climate change issues and a promoter of the international agreement concluded in 1997 in Kyoto, Japan. Mr. Clapp continued to attend subsequent global warming talks even after the US Congress did not ratify the Kyoto accord.'

[10] John Holdren has defended the use of the phrase 'Research Center' since research is carried out with sponsorship by National Science Foundation, the National Oceanographic Administration, and NASA. However, it is hardly uncommon to find sponsorship of the activities of environmental NGO's by federal funding agencies.

*Climate Science: Is it currently designed to answer questions?*

the bulk of the educated public is unable to follow scientific arguments; 'knowing' that all scientists agree relieves them of any need to do so. Second, such a claim serves as a warning to scientists that the topic at issue is a bit of a minefield that they would do well to avoid.

The myth of scientific consensus is also perpetuated in the web's Wikipedia where climate articles are vetted by William Connolley, who regularly runs for office in England as a Green Party candidate. No deviation from the politically correct line is permitted.

Perhaps the most impressive exploitation of climate science for political purposes has been the creation of the Intergovernmental Panel on Climate Change (IPCC) by two UN agencies, UNEP (United Nations Environmental Program) and WMO (World Meteorological Organization), and the agreement of all major countries at the 1992 Rio Conference to accept the IPCC as authoritative. Formally, the IPCC summarizes the peer reviewed literature on climate every five years. On the face of it, this is an innocent and straightforward task. One might reasonably wonder why it takes 100's of scientists five years of constant travelling throughout the world in order to perform this task. The charge to the IPCC is not simply to summarize, but rather to provide the science with which to support the negotiating process whose aim is to control greenhouse gas levels. This is a political rather than a scientific charge. That said, the participating scientists have some leeway in which to reasonably describe matters, since the primary document that the public associates with the IPCC is not the extensive report prepared by the scientists, but rather the Summary for Policymakers which is written by an assemblage of representative from governments and NGO's, with only a small scientific representation[11],[12].

---

[11] Appendix 1 is the invitation to the planning session for the 5$^{th}$ assessment. It clearly emphasizes strengthening rather than checking the IPCC position. Appendix 2 reproduces a commentary by Stephen McIntyre on the recent OfCom findings concerning a British TV program opposing global warming alarmism. The response of the IPCC officials makes it eminently clear that the IPCC is fundamentally a political body. If further evidence were needed, one simply has to observe the fact that the IPCC Summary for Policymakers will selectively cite results to emphasize negative consequences. Thus the summary for Working Group II observes that global warming will result in "Hundreds of millions of people exposed to increased water stress." This, however, is based on work (Arnell, 2004) which actually shows that by the 2080s the net global population at risk declines by up to 2.1 billion people (depending on which scenario one wants to emphasize)! The IPCC further ignores the capacity to build reservoirs to alleviate those areas they project as subject to drought (I am indebted to Indur Goklany for noting this example.)

[12] Appendix 3 is a recent op-ed from the Boston Globe, written by the aforementioned John Holdren. What is interesting about this piece is that what little science it invokes is overtly incorrect. Rather, it points to the success of the above process of taking over scientific institutions as evidence of the correctness of global warming alarmism. The 3 atmospheric scientists who are explicitly mentioned are chemists with no particular expertise in climate, itself. While, Holdren makes much of the importance of expertise, he fails to note that he, himself, is hardly a contributor to the science of climate. Holdren and Paul Ehrlich (of *Population Bomb* fame; in that work he predicted famine and food riots for the US in the 1980's)

*Climate Science: Is it currently designed to answer questions?*

## 3. Science in the service of politics

Given the above, it would not be surprising if working scientists would make special efforts to support the global warming hypothesis. There is ample evidence that this is happening on a large scale. A few examples will illustrate this situation. Data that challenges the hypothesis are simply changed. In some instances, data that was thought to support the hypothesis is found not to, and is then changed. The changes are sometimes quite blatant, but more often are somewhat more subtle. The crucial point is that geophysical data is almost always at least somewhat uncertain, and methodological errors are constantly being discovered. Bias can be introduced by simply considering only those errors that change answers in the desired direction. The desired direction in the case of climate is to bring the data into agreement with models, even though the models have displayed minimal skill in explaining or predicting climate. Model projections, it should be recalled, are the basis for our greenhouse concerns. That corrections to climate data should be called for, is not at all surprising, but that such corrections should always be in the 'needed' direction is exceedingly unlikely. Although the situation suggests overt dishonesty, it is entirely possible, in today's scientific environment, that many scientists feel that it is the role of science to vindicate the greenhouse paradigm for climate change as well as the credibility of models. Comparisons of models with data are, for example, referred to as model validation studies rather than model tests.

The first two examples involve paleoclimate simulations and reconstructions. Here, the purpose has been to show that both the models and the greenhouse paradigm can explain past climate regimes, thus lending confidence to the use of both to anticipate future changes. In both cases (the Eocene about 50 million years ago, and the Last Glacial Maximum about 18 thousand years ago), the original data were in conflict with the greenhouse paradigm as implemented in current models, and in both cases, lengthy efforts were made to bring the data into agreement with the models.

In the first example, the original data analysis for the Eocene (Shackleton and Boersma, 1981) showed the polar regions to have been so much warmer than the present that a type of alligator existed on Spitzbergen as did florae and fauna in Minnesota that could not have survived frosts. At the same time, however, equatorial temperatures were found to be about 4K colder than at present. The first attempts to simulate the Eocene (Barron, 1987) assumed that the warming would be due to high levels of $CO_2$, and using a climate GCM (General Circulation Model), he obtained relatively uniform warming at all latitudes, with the meridional gradients remaining

---

are responsible for the I=PAT formula. Holdren, somewhat disingenuously claims that this is merely a mathematical identity where I is environmental impact, P is population, A is GDP/P and T is I/GDP. However, in popular usage, A has become affluence and T has become technology (viz Schneider, 1997; see also *Wikipedia*).

## *Climate Science: Is it currently designed to answer questions?*

much as they are today.  This behavior continues to be the case with current GCMs (Huber, 2008).  As a result, paleoclimatologists have devoted much effort to 'correcting' their data, but, until very recently, they were unable to bring temperatures at the equator higher than today's (Schrag, 1999, Pearson et al, 2000).  However, the latest paper (Huber, 2008) suggests that the equatorial data no longer constrains equatorial temperatures at all, and any values may have existed.  All of this is quite remarkable since there is now evidence that current meridional distributions of temperature depend critically on the presence of ice, and that the model behavior results from improper tuning wherein present distributions remain even when ice is absent.

The second example begins with the results of a major attempt to observationally reconstruct the global climate of the last glacial maximum (CLIMAP, 1976).  Here it was found that although extratropical temperatures were much colder, equatorial temperatures were little different from today's.  There were immediate attempts to simulate this climate with GCMs and reduced levels of $CO_2$.  Once again the result was lower temperatures at all latitudes (Bush and Philander, 1998a,b), and once again, numerous efforts were made to 'correct' the data.  After much argument, the current position appears to be that tropical temperatures may have been a couple of degrees cooler than today's.  However, papers appeared claiming far lower temperatures (Crowley, 2000).  We will deal further with this issue in the next section where we describe papers that show that the climate associated with ice ages is well described by the Milankovich Hypothesis that does not call for any role for $CO_2$.

Both of the above examples probably involved legitimate corrections, but only corrections that sought to bring observations into agreement with models were initially considered, thus avoiding the creative conflict between theory and data that has characterized the past successes of science.  To be sure, however, the case of the Last Glacial Maximum shows that climate science still retains a capacity for self-correction.

The next example has achieved a much higher degree of notoriety than the previous two.  In the first IPCC assessment (IPCC, 1990), the traditional picture of the climate of the past 1100 years was presented.  In this picture, there was a medieval warm period that was somewhat warmer than the present as well as the little ice age that was cooler.  The presence of a period warmer than the present in the absence of any anthropogenic greenhouse gases was deemed an embarrassment for those holding that present warming could only be accounted for by the activities of man.  Not surprisingly, efforts were made to get rid of the medieval warm period (According to Demming, 2005,  in 1995, "A major person working in the area of climate change and global warming sent me an astonishing email that said "We have to get rid of the Medieval Warm Period.""  Although Deming did not name the individual because he could not locate the email, he did note in an email to me that "Off the record, and over the years, I may have confided verbally to a few persons what my recollection was of the person's identity." That such attitudes among climate scientists exists is evident, for example, in the released emails from the University of East Anglia (sometimes referred to as 'climategate') where Jonathan Overpeck is chastised for speaking of 'nailing the MWP (Medieval Warm Period)' (http://foia2011.org/index.php?id=212)  This is, indeed, a questionable position for someone

*Climate Science: Is it currently designed to answer questions?*

who served as an organizer for the IPCC Fifth Assessment Report (viz Appendix 1).).

The most infamous effort was that due to Mann et al (1998, 1999[13]) which used primarily a few handfuls of tree ring records to obtain a reconstruction of Northern Hemisphere temperature going back eventually a thousand years that no longer showed a medieval warm period. Indeed, it showed a slight cooling for almost a thousand years culminating in a sharp warming beginning in the nineteenth century. The curve came to be known as the hockey stick, and featured prominently in the next IPCC report, where it was then suggested that the present warming was unprecedented in the past 1000 years. The study immediately encountered severe questions concerning both the proxy data and its statistical analysis (interestingly, the most penetrating critiques came from outside the field: McIntyre and McKitrick, 2003, 2005a,b). This led to two independent assessments of the hockey stick (Wegman,2006, North, 2006), both of which found the statistics inadequate for the claims. The story is given in detail in Holland (2007). Since the existence of a medieval warm period is amply documented in historical accounts for the North Atlantic region (Soon et al, 2003), Mann et al countered that the warming had to be regional but not characteristic of the whole northern hemisphere. Given that an underlying assumption of their analysis was that the geographic pattern of warming had to have remained constant, this would have invalidated the analysis *ab initio* without reference to the specifics of the statistics. Indeed, the 4[th] IPCC (IPCC, 2007) assessment no longer featured the hockey stick, but the claim that current warming is unprecedented remains, and Mann et al's reconstruction is still shown in Chapter 6 of the 4[th] IPCC assessment, buried among other reconstructions. Here too, we will return to this matter briefly in the next section.

The fourth example is perhaps the strangest. For many years, the global mean temperature record showed cooling from about 1940 until the early 70's. This, in fact, led to the concern for global cooling during the 1970's. The IPCC regularly, through the 4[th] assessment, boasted of the ability of models to simulate this cooling (while failing to emphasize that each model required a different specification of completely undetermined aerosol cooling in order to achieve this simulation (Kiehl, 2007)). Improvements in our understanding of aerosols are increasingly making such arbitrary tuning somewhat embarrassing, and, no longer surprisingly, the data has been 'corrected' to get rid of the mid 20[th] century cooling (Thompson et al, 2008). This may, in fact, be a legitimate correction (http://www.climateaudit.org/?p=3114). The embarrassment may lie in the continuous claims of modelers to have simulated the allegedly incorrect data.

The fifth example deals with the fingerprint of warming. It has long been noted that greenhouse warming is primarily centered in the upper troposphere (Lindzen, 1999) and, indeed, model's show that the maximum rate of warming is found in the upper tropical troposphere (Lee, et al, 2007). Lindzen (2007) noted that temperature data from both satellites and balloons failed to show such a maximum. This, in turn, permitted one to bound the greenhouse contribution to surface warming, and led to an estimate of climate sensitivity that was appreciably less than

---

[13] The 1998 paper actually only goes back to 1400 CE, and acknowledges that there is no useful resolution of spatial patterns of variability going further back. It is the 1999 paper that then goes back 1000 years.

*Climate Science: Is it currently designed to answer questions?*

found in current models. Once the implications of the observations were clearly identified, it was only a matter of time before the data were 'corrected.' The first attempt came quickly (Vinnikov et al, 2006) wherein the satellite data was reworked to show large warming in the upper troposphere, but the methodology was too blatant for the paper to be commonly cited[14]. There followed an attempt wherein the temperature data was rejected, and where temperature trends were inferred from wind data (Allen and Sherwood, 2008). Over sufficiently long periods, there is a balance between vertical wind shear and meridional temperature gradients (the thermal wind balance), and, with various assumptions concerning boundary conditions, one can, indeed, infer temperature trends, but the process involves a more complex, indirect, and uncertain procedure than is involved in directly measuring temperature. Moreover, as Pielke et al (2008) have noted, the results display a variety of inconsistencies. They are nonetheless held to resolve the discrepancy with models.

The sixth example takes us into astrophysics. Since the 1970's, considerable attention has been given to something known as the Early Faint Sun Paradox. This paradox was first publicized by Sagan and Mullen (1972). They noted that the standard model for the sun robustly required that the sun brighten with time so that 2-3 billion years ago, it was about 30% dimmer than it is today (recall that a doubling of $CO_2$ corresponds to only a 2% change in the radiative budget). One would have expected that the earth would have been frozen over, but the geological evidence suggested that the ocean was unfrozen. Attempts were made to account for this by an enhanced greenhouse effect. Sagan and Mullen (1972) suggested an ammonia rich atmosphere might work. Others suggested an atmosphere with as much as several bars of $CO_2$ (recall that currently $CO_2$ is about 380 parts per million of a 1 bar atmosphere). Finally, Kasting and colleagues tried to resolve the paradox with large amounts of methane. For a variety of reasons, all these efforts were deemed inadequate[15] (Haqqmisra et al, 2008). There followed a remarkable attempt to get rid of the standard model of the sun (Sackman and Boothroyd, 2003). This is not exactly the same as altering the data, but the spirit is the same. The paper claimed to have gotten rid of the paradox. However, in fact, the altered model still calls for substantial brightening, and, moreover, does not seem to have gotten much acceptance among solar modelers.

My last specific example involves the social sciences. Given that it has been maintained since at least 1988 that all scientists agree about alarming global warming, it is embarrassing to have scientists objecting to the alarm. To 'settle' the matter, a certain Naomi Oreskes published a paper in *Science* (Oreskes, 2004) purporting to have surveyed the literature and not have found a single paper questioning the alarm (Al Gore offers this study as proof of his own correctness in "Inconvenient Truth."). Both Benny Peiser (a British sociologist) and Dennis Bray (an historian of science) noted obvious methodological errors, but *Science* refused to publish these rebuttals

---

[14] Of course, Vinnikov et al did mention it. When I gave a lecture at Rutgers University in October 2007, Alan Robock, a professor at Rutgers and a coauthor of Vinnikov et al declared that the 'latest data' resolved the discrepancy wherein the model fingerprint could not be found in the data.

[15] Haqqmisra, a graduate student at the Pennsylvania State University, is apparently still seeking greenhouse solutions to the paradox.

*Climate Science: Is it currently designed to answer questions?*

with no regard for the technical merits of the criticisms presented [16]. Moreover, Oreskes was a featured speaker at the celebration of Spencer Weart's thirty years as head of the American Institute of Physics' Center for History of Physics. Weart, himself, had written a history of the global warming issue (Weart, 2003) where he repeated, without checking, the slander taken from a screed by Ross Gelbspan (*The Heat is On*) in which I was accused of being a tool of the fossil fuel industry. Weart also writes with glowing approval of Gore's *Inconvenient Truth*. As far as Oreskes' claim goes, it is clearly absurd[17]. A more carefully done study revealed a very different picture (Schulte, 2007)

The above examples do not include the most convenient means whereby nominal scientists can support global warming alarm: namely, the matter of impacts. Here, scientists who generally have no knowledge of climate physics at all, are supported to assume the worst projections of global warming and imaginatively suggest the implications of such warming for whatever field they happen to be working in. This has led to the bizarre claims that global warming will contribute to kidney stones, obesity, cockroaches, noxious weeds, sexual imbalance in fish, etc. The scientists who participate in such exercises quite naturally are supportive of the catastrophic global warming hypothesis despite their ignorance of the underlying science[18].

---

[16] The refusal was not altogether surprising. The editor of *Science*, at the time, was Donald Kennedy, a biologist (and colleague of Paul Ehrlich and Stephen Schneider, both also members of Stanford's biology department), who had served as president of Stanford University. His term, as president, ended with his involvement in fiscal irregularities such as charging to research overhead such expenses as the maintenance of the presidential yacht and the provision of flowers for his daughter's wedding – offering peculiar evidence for the importance of grant overhead to administrators. Kennedy had editorially declared that the debate concerning global warming was over and that skeptical articles would not be considered. More recently, he has published a relatively pure example of Orwellian double-speak (Kennedy, 2008) wherein he called for better media coverage of global warming, where by 'better' he meant more carefully ignoring any questions about global warming alarm. As one might expect, Kennedy made extensive use of Oreskes' paper. He also made the remarkably dishonest claim that the IPCC Summary for Policymakers was much more conservative than the scientific text.

[17] Oreskes, apart from overt errors, merely considered support to consist in agreement that there had been *some* warming, and that anthropogenic $CO_2$ contributed *part* of the warming. Such innocent conclusions have essentially nothing to do with catastrophic projections. Moreover, most of the papers she looked at didn't even address these issues; they simply didn't question these conclusions.

[18] Perhaps unsurprisingly, The Potsdam Institute, home of Greenpeace's Bill Hare, now has a Potsdam Institute for Climate Impact Research.

*Climate Science: Is it currently designed to answer questions?*

## 4. Pressures to inhibit inquiry and problem solving

It is often argued that in science the truth must eventually emerge. This may well be true, but, so far, attempts to deal with the science of climate change objectively have been largely forced to conceal such truths as may call into question global warming alarmism (even if only implicitly). The usual vehicle is peer review, and the changes imposed were often made in order to get a given paper published. Publication is, of course, essential for funding, promotion, etc. The following examples are but a few from cases that I am personally familiar with. These, almost certainly, barely scratch the surface. What is generally involved, is simply the inclusion of an irrelevant comment supporting global warming accepted wisdom. When the substance of the paper is described, it is generally claimed that the added comment represents the 'true' intent of the paper. In addition to the following examples, Appendix 2 offers excellent examples of 'spin control.'.

As mentioned in the previous section, one of the reports assessing the Mann et al Hockey Stick was prepared by a committee of the US National Research Counsel (a branch of the National Academy) chaired by Gerald North (North, 2006). The report concluded that the analysis used was totally unreliable for periods longer ago than about 400 years. In point of fact, the only basis for the 400 year choice was that this brought one to the midst of the Little Ice Age, and there is essentially nothing surprising about a conclusion that we are now warmer. Still, without any basis at all, the report also concluded that despite the inadequacy of the Mann et al analysis, the conclusion might still be correct. It was this baseless conjecture that received most of the publicity surrounding the report.

In a recent paper, Roe (2006) showed that the orbital variations in high latitude summer insolation correlate excellently with changes in glaciation – once one relates the insolation properly to the rate of change of glaciation rather than to the glaciation itself. This provided excellent support for the Milankovich hypothesis. Nothing in the brief paper suggested the need for any other mechanism. Nonetheless, Roe apparently felt compelled to include an irrelevant caveat stating that the paper had no intention of ruling out a role for $CO_2$.

Choi and Ho (2006, 2008) published interesting papers on the optical properties of high tropical cirrus that largely confirmed earlier results by Lindzen, Chou and Hou (2001) on an important negative feedback (the iris effect – something that we will describe later in this section) that would greatly reduce the sensitivity of climate to increasing greenhouse gases. A proper comparison required that the results be normalized by a measure of total convective activity, and, indeed, such a comparison was made in the original version of Choi and Ho's paper. However, reviewers insisted that the normalization be removed from the final version of the paper which left the relationship to the earlier paper unclear.

Horvath and Soden (2008) found observational confirmation of many aspects of the iris effect, but accompanied these results with a repetition of criticisms of the iris effect that were irrelevant and even contradictory to their own paper. The point, apparently, was to suggest that despite

*Climate Science: Is it currently designed to answer questions?*

their findings, there might be other reasons to discard the iris effect. Later in this section, I will return to these criticisms. However, the situation is far from unique. I have received preprints of papers wherein support for the iris was found, but where this was omitted in the published version of the papers

In another example, I had originally submitted a paper mentioned in the previous section (Lindzen, 2007) to *American Scientist*, the periodical of the scientific honorary society in the US, Sigma Xi, at the recommendation of a former officer of that society. There followed a year of discussions, with an editor, David Schneider, insisting that I find a coauthor who would illustrate why my paper was wrong. He argued that publishing something that contradicted the IPCC was equivalent to publishing a paper that claimed that 'Einstein's general theory of relativity is bunk.' I suggested that it would be more appropriate for *American Scientist* to solicit a separate paper taking a view opposed to mine. This was unacceptable to Schneider, so I ended up publishing the paper elsewhere. Needless to add, Schneider has no background in climate physics. At the same time, a committee consisting almost entirely in environmental activists led by Peter Raven, the ubiquitous John Holdren, Richard Moss, Michael MacCracken, and Rosina Bierbaum were issuing a joint Sigma Xi - United Nations Foundation (the latter headed by former Senator and former Undersecretary of State Tim Wirth[19] and founded by Ted Turner) report endorsing global warming alarm, to a degree going far beyond the latest IPCC report. I should add that simple disagreement with conclusions of the IPCC has become a common basis for rejecting papers for publication in professional journals – as long as the disagreement suggests reduced alarm. An example will be presented later in this section.

Despite all the posturing about global warming, more and more people are becoming aware of the fact that global mean temperatures have not increased statistically significantly since 1995. One need only look at the temperature records posted on the web by the Hadley Centre. The way this is acknowledged in the literature forms a good example of the spin that is currently required to maintain global warming alarm. Recall that the major claim of the IPCC 4$^{th}$ Assessment was that there was a 90% certainty that most of the warming of the preceding 50 years was due to man (whatever that might mean). This required the assumption that what is known as natural internal variability (ie, the variability that exists without any external forcing and represents the fact that the climate system is never in equilibrium) is adequately handled by the existing climate models. The absence of any net global warming over the last dozen years or so, suggests that this assumption may be wrong. Smith et al (2007) (Smith is with the Hadley Centre in the UK) acknowledged this by pointing out that initial conditions had to reflect the disequilibrium at some starting date, and when these conditions were 'correctly' chosen, it was possible to better replicate the period without warming. This acknowledgment of error was

---

[19] Tim Wirth chaired the hearing where Jim Hansen rolled out the alleged global warming relation to the hot summer of 1988 (viz Section 2). He is noted for having arranged for the hearing room to have open windows to let in the heat so that Hansen would be seen to be sweating for the television cameras. Wirth is also frequently quoted as having said "We've got to ride the global warming issue. Even if the theory of global warming is wrong, we will be doing the right thing — in terms of economic policy and environmental policy."

*Climate Science: Is it currently designed to answer questions?*

accompanied by the totally unjustified assertion that global warming would resume with a vengeance in 2009[20]. As 2009 approaches and the vengeful warming seems less likely to occur, a new paper came out (this time from the Max Planck Institute: Keenlyside et al, 2008) moving the date for anticipated resumption of warming to 2015. It is indeed a remarkable step backwards for science to consider models that have failed to predict the observed behavior of the climate to nonetheless have the same validity as the data[21].

Tim Palmer, a prominent atmospheric scientist at the European Centre for Medium Range Weather Forecasting, is quoted by Fred Pearce (Pearce, 2008) in the *New Scientist* as follows: "Politicians seem to think that the science is a done deal," says Tim Palmer. "I don't want to undermine the IPCC, but the forecasts, especially for regional climate change, are immensely uncertain." Pearce, however, continues "Palmer .. does not doubt that the Intergovernmental Panel on Climate Change (IPCC) has done a good job alerting the world to the problem of global climate change. But he and his fellow climate scientists are acutely aware that the IPCC's predictions of how the global change will affect local climates are little more than guesswork. They fear that if the IPCC's predictions turn out to be wrong, it will provoke a crisis in confidence that undermines the whole climate change debate. On top of this, some climate scientists believe that even the IPCC's global forecasts leave much to be desired. ..." Normally, one would think that undermining the credibility of something that is wrong is appropriate.

Even in the present unhealthy state of science, papers that are overtly contradictory to the catastrophic warming scenario do get published (though not without generally being substantially watered down during the review process). They are then often subject to the remarkable process of 'discreditation.' This process consists in immediately soliciting attack papers that are published quickly as independent articles rather than comments. The importance of this procedure is as follows. Normally such criticisms are published as comments, and the original authors are able to respond immediately following the comment. Both the comment and reply are published together. By publishing the criticism as an article, the reply is published as a correspondence, which is usually delayed by several months, and the critics are permitted an immediate reply. As a rule, the reply of the original authors is ignored in subsequent references.

In 2001, I published a paper (Lindzen, Chou and Hou) that used geostationary satellite data to suggest the existence of a strong negative feedback that we referred to as the Iris Effect. The gist

---

[20] When I referred to the Smith et al paper at a hearing of the European Parliament, Professor Schellnhuber of the Potsdam Institute (which I mentioned in the previous section with respect to its connection to Greenpeace) loudly protested that I was being 'dishonest' by not emphasizing what he referred to as the main point in Smith et al: namely that global warming would return with a vengeance.

[21] The matter of 'spin control' warrants a paper by itself. In connection with the absence of warming over the past 13 years, the common response is that 7 of the last 10 warmest years in the record occurred during the past decade. This is actually to be expected, given that we are in a warm period, and the temperature is always fluctuating. However, it has nothing to do with trends.

*Climate Science: Is it currently designed to answer questions?*

of the feedback is that upper level stratiform clouds in the tropics arise by detrainment from cumulonimbus towers, that the radiative impact of the stratiform clouds is primarily in the infrared where they serve as powerful greenhouse components, and that the extent of the detrainment decreases markedly with increased surface temperature. The negative feedback resulted from the fact that the greenhouse warming due to the stratiform clouds diminished as the surface temperature increased, and increased as the surface temperature decreased – thus resisting the changes in surface temperature. The impact of the observed effect was sufficient to greatly reduce the model sensitivities to increasing $CO_2$, and it was, moreover, shown that models failed to display the observed cloud behavior. The paper received an unusually intense review from four reviewers. Once the paper appeared, the peer review editor of the *Bulletin of the American Meteorological Society*, Irwin Abrams, was replaced by a new editor, Jeffrey Rosenfeld (holding the newly created position of Editor in Chief), and the new editor almost immediately accepted a paper criticizing our paper (Hartmann and Michelsen, 2002), publishing it as a separate paper rather than a response to our paper (which would have been the usual and appropriate procedure). In the usual procedure, the original authors are permitted to respond in the same issue. In the present case, the response was delayed by several months. Moreover, the new editor chose to label the criticism as follows: "Careful analysis of data reveals no shrinkage of tropical cloud anvil area with increasing SST." In fact, this criticism was easily dismissed. The claim of Hartmann and Michelsen was that the effect we observed was due to the intrusion of midlatitude non-convective clouds into the tropics. If this were true, then the effect should have diminished as one restricted observations more closely to the equator, but as we showed (Lindzen, Chou and Hou, 2002), exactly the opposite was found. There were also separately published papers (again violating normal protocols allowing for immediate response) by Lin et al, 2002 and Fu, Baker and Hartmann, 2002, that criticized our paper by claiming that since the instruments on the geostationary satellite could not see the thin stratiform clouds that formed the tails of the clouds we could see, that we were not entitled to assume that the tails existed. Without the tails, the radiative impact of the clouds would be primarily in the visible where the behavior we observed would lead to a positive feedback; with the tails the effect is a negative feedback. The tails had long been observed, and the notion that they abruptly disappeared when not observed by an insufficiently sensitive sensor was absurd on the face of it, and the use of better instruments by Choi and Ho (2006, 2008) confirmed the robustness of the tails and the strong dominance of the infrared impact. However, as we have already seen, virtually any mention of the iris effect tends to be accompanied with a reference to the criticisms, a claim that the theory is 'discredited,' and absolutely no mention of the responses. This is even required of papers that are actually supporting the iris effect.

Vincent Courtillot et al (2007) encountered a similar problem. (Courtillot, it should be noted, is the director of the Institute for the Study of the Globe at the University of Paris.) They found that time series for magnetic field variations appeared to correlate well with temperature measurements – suggesting a possible non-anthropogenic source of forcing. This was immediately criticized by Bard and Delaygue (2008), and Courtillot et al were given the conventional right to reply which they did in a reasonably convincing manner. What followed, however, was highly unusual. Raymond Pierrehumbert (a professor of meteorology at the

*Climate Science: Is it currently designed to answer questions?*

University of Chicago and a fanatical environmentalist) posted a blog supporting Bard and Delaygue, accusing Courtillot et al of fraud, and worse. Alan Robock (a coauthor of Vinnikov et al mentioned in the preceding section) perpetuated the slander in a letter circulated to all officers of the American Geophysical Union. The matter was then taken up (in December of 2007) by major French newspapers (*LeMonde*, *Liberation*, and *Le Figaro*) that treated Pierrehumbert's defamation as fact. As in the previous case, all references to the work of Courtillot et al refer to it as 'discredited' and no mention is made of their response. Moreover, a major argument against the position of Courtillot et al is that it contradicted the claim of the IPCC.

In 2005, I was invited by Erneso Zedillo to give a paper at a symposium he was organizing at his Center for Sustainability Studies at Yale. The stated topic of the symposium was Global Warming Policy After 2012, and the proceedings were to appear in a book to by entitled Global Warming: Looking Beyond Kyoto. Only two papers dealing with global warming science were presented: mine and one by Stefan Rahmstorf of the Potsdam Institute. The remaining papers all essentially assumed an alarming scenario and proceeded to discuss economics, impacts, and policy. Rahmstorf and I took opposing positions, but there was no exchange at the meeting, and Rahmstorf had to run off to another meeting. As agreed, I submitted the manuscript of my talk, but publication was interminably delayed, perhaps because of the presence of my paper. In any event, the Brookings Institute (a centrist Democratic Party think tank) agreed to publish the volume. When the volume finally appeared (Zedillo, 2008), I was somewhat shocked to see that Rahmstorf's paper had been modified from what he presented, and had been turned into an attack not only on my paper but on me personally[22]. I had received no warning of this; nor was I given any opportunity to reply. Inquiries to the editor and the publisher went unanswered. Moreover, the Rahmstorf paper was moved so that it immediately followed my paper. The reader is welcome to get a copy of the exchange, including my response, on my web site (Lindzen-Rahmstorf Exchange, 2008), and judge the exchange for himself.

One of the more bizarre tools of global warming revisionism is the posthumous alteration of skeptical positions.

Thus, the recent deaths of two active and professionally prominent skeptics, Robert Jastrow (the founding director of NASA's Goddard Institute for Space Studies, now headed by James Hansen), and Reid Bryson (a well known climatologist at the University of Wisconsin) were accompanied by obituaries suggesting deathbed conversions to global warming alarm.

The death of another active and prominent skeptic, William Nierenberg (former director of the Scripps Oceanographic Institute), led to the creation of a Nierenberg Prize that is annually awarded to an environmental activist. The most recent recipient was James Hansen who Nierenberg detested.

---

[22] The strange identification of the $CO_2$ caused global warming paradigm with general relativity theory, mentioned earlier in this section, is repeated by Rahmstorf. This repetition of odd claims may be a consequence of the networking described in footnote 7.

*Climate Science: Is it currently designed to answer questions?*

Perhaps the most extraordinary example of this phenomenon involves a paper by Singer, Starr, and Revelle (1991). In this paper, it was concluded that we knew too little about climate to implement any drastic measures. Revelle, it may be recalled, was the professor that Gore credits with introducing him to the horrors of $CO_2$ induced warming. There followed an intense effort led by a research associate at Harvard, Justin Lancaster, in coordination with Gore staffers, to have Revelle's name posthumously removed from the published paper. It was claimed that Singer had pressured an old and incompetent man to allow his name to be used. To be sure, everyone who knew Revelle, felt that he had been alert until his death. There followed a law suit by Singer, where the court found in Singer's favor. The matter is described in detail in Singer (2003).

Occasionally, prominent individual scientists do publicly express skepticism. The means for silencing them are fairly straightforward.

Will Happer, director of research at the Department of Energy (and a professor of physics at Princeton University) was simply fired from his government position after expressing doubts about environmental issues in general. His case is described in Happer (2003).

Michael Griffin, NASA's administrator, publicly expressed reservations concerning global warming alarm in 2007. This was followed by a barrage of ad hominem attacks from individuals including James Hansen and Michael Oppenheimer. Griffin has since stopped making any public statements on this matter.

Freeman Dyson, an acknowledged great in theoretical physics, managed to publish a piece in *New York Review of Books* (Dyson, 2008), where in the course of reviewing books by Nordhaus and Zedillo (the latter having been referred to earlier), he expressed cautious support for the existence of substantial doubt concerning global warming. This was followed by a series of angry letters as well as condemnation on the realclimate.org web site including ad hominem attacks. Given that Dyson is retired, however, there seems little more that global warming enthusiasts can do. However, we may hear of a deathbed conversion in the future.

## 5. Dangers for science and society

This paper has attempted to show how changes in the structure of scientific activity over the past half century have led to extreme vulnerability to political manipulation. In the case of climate change, these vulnerabilities have been exploited to a remarkable extent. The dangers that the above situation poses for both science and society are too numerous to be discussed in any sort of adequate way in this paper. It should be stressed that the climate change issue, itself, constitutes a major example of the dangers intrinsic to the structural changes in science.

As concerns the specific dangers pertaining to the climate change issue, we are already seeing that the tentative policy moves associated with 'climate mitigation' are contributing to deforestation, food riots, potential trade wars, inflation, energy speculation and overt corruption

## Climate Science: Is it currently designed to answer questions?

as in the case of ENRON (one of the leading lobbyists for Kyoto prior to its collapse). There is little question that global warming has been exploited many governments and corporations (and not just by ENRON; Lehman Brothers, for example, was also heavily promoting global warming alarm, and relying on the advice of James Hansen, etc.) for their own purposes, but it is unclear to what extent such exploitation has played an initiating role in the issue. The developing world has come to realize that the proposed measures endanger their legitimate hopes to escape poverty, and, in the case of India, they have, encouragingly, led to an assessment of climate issues independent of the 'official' wisdom (Government of India, 2008[23]). For purposes of this paper, however, I simply want to briefly note the specific implications for science and its interaction with society. Although society is undoubtedly aware of the imperfections of science, it has rarely encountered a situation such as the current global warming hysteria where institutional science has so thoroughly committed itself to policies which call for massive sacrifices in well being world wide. Past scientific errors did not lead the public to discard the view that science on the whole was a valuable effort. However, the extraordinarily shallow basis for the commitment to climate catastrophe, and the widespread tendency of scientists to use unscientific means to arouse the public's concerns, is becoming increasingly evident, and the result could be a reversal of the trust that arose from the triumphs of science and technology during the World War II period. Further, the reliance by the scientific community on fear as a basis for support, may, indeed, have severely degraded the ability of science to usefully address problems that need addressing. It should also be noted that not all the lessons of the World War II period have been positive. Massive crash programs such as the Manhattan Project are not appropriate to all scientific problems. In particular, such programs are unlikely to be effective in fields where the basic science is not yet in place. Rather, they are best suited to problems where the needs are primarily in the realm of engineering.

Although the change in scientific culture has played an important role in making science more vulnerable to exploitation by politics, the resolution of specific issues may be possible without explicitly addressing the structural problems in science. In the US, where global warming has become enmeshed in partisan politics, there is a natural opposition to exploitation which is not specifically based on science itself. However, the restoration of the traditional scientific paradigm will call for more serious efforts. Such changes are unlikely to come from any fiat. Nor is it likely to be implemented by the large science bureaucracies that have helped create the problem in the first place. A potentially effective approach would be to change the incentive structure of science. The current support mechanisms for science is one where the solution of a scientific problem is rewarded by ending support. This hardly encourages the solution of problems or the search for actual answers. Nor does it encourage meaningfully testing hypotheses. The alternative calls for a measure of societal trust, patience, and commitment to elitism that hardly seems consonant with the contemporary attitudes. It may, however, be possible to make a significant beginning by carefully reducing the funding for science. Many

---

[23] A curious aspect of the profoundly unalarming Indian report is the prominent involvement in the preparation of the report by Dr. Rajendra Pachauri (an economist and long term UN bureaucrat) who heads the IPCC. Dr. Pachauri has recently been urging westerners to reduce meat consumption in order to save the earth from destruction by global warming.

*Climate Science: Is it currently designed to answer questions?*

scientists would be willing to accept a lower level of funding in return for greater freedom and stability. Other scientists may find the trade-off unacceptable and drop out of the enterprise. The result, over a period of time, could be a gradual restoration of a better incentive structure. One ought not underestimate the institutional resistance to such changes, but the alternatives are proving to be much worse. Some years ago, I described some of what I have discussed here at a meeting in Erice (Lindzen, 2005). Richard Garwin (who some regard as the inventor of the H-bomb) rose indignantly to state that he did not want to hear such things. Quite frankly, I also don't want to hear such things. However, I fear that ignoring such things will hardly constitute a solution, and a solution may be necessary for the sake of the scientific enterprise.

**Acknowledgments.** The author wishes to thank Dennis Ambler, Willie Soon, Lubos Motl and Nigel Lawson for useful comments and assistance.

# Appendix 1

July 11, 2008

On behalf of the organizing committee, and workshop co-sponsors IPCC, WCRP, IGBP, the US National Science Foundation, and Climate Central, we take great pleasure in inviting you to attend a **"Joint IPCC-WCRP-IGBP Workshop: New Science Directions and Activities Relevant to the IPCC AR5"** to be held **March 3—6, 2009**. The Workshop will be hosted by the International Pacific Research Center (IPRC) at the University of Hawaii in Honolulu, Hawaii. The workshop is open to WG1 LAs and CLAs from all four assessments. The proceedings will be made available to IPCC.

This workshop has several major goals:
1) New science results and research directions relevant for the upcoming IPCC Fifth Assessment Report (AR5) will be discussed, with a view to the manner in which new observations and models can ensure their fullest possible consideration in the upcoming AR5. This could include but are not limited to e.g., ice sheet instability, land use parameterizations, aerosols and their effects on clouds and climate, new attribution results beyond temperature, and improved ENSO projections.

2) Subsequent to the AR4, an international planning process has begun to perform a coordinated set of climate model experiments with AOGCMs as well as emerging Earth System Models (ESMs, including new aspects of climate-vegetation and carbon cycle feedbacks) to quantify time-evolving regional climate change using mitigation/adaptation scenarios. These experiments will address key feedbacks in climate system response to increasing greenhouse gases. For example, carbon cycle feedback was identified as one of the main uncertainties for the upper end of future climate projections in the AR4. An international process to produce a set of mitigation scenarios for use in WG1, termed Representative Concentration Pathways (RCPs), will culminate in the fall of 2008 when the scenarios will be turned over to the WG1

*Climate Science: Is it currently designed to answer questions?*

modeling groups. The ingredients in these scenarios (emissions and concentrations of various constituents) will be reviewed at the workshop to ensure they are compatible with what is required by the new Earth System Models. It is essential that scientists gathered at the workshop examine and discuss them in detail to ensure compatibility and consistency with the new ESMs, particularly with regard to land use/land cover and emissions, which will also be a central topic at the workshop. Additionally, output requirements for the model simulations and a strategy for extension of long-term simulations to 2300 will be discussed.

3) Decadal climate prediction has recently emerged as a research activity that combines aspects of seasonal/interannual predictions and longer term emission scenario-driven climate change. Recent research results, as well as plans for coordinated experiments to address science problems associated with the decadal prediction, will be discussed at the workshop.

For planning purposes, please **register for the workshop at** http://www.regonline.com/Checkin.asp?EventId=633780 before September 1, 2008. Hotel information is available on that web site, and participants are encouraged to make their hotel reservations as soon as possible because reservations for the various hotel options are on a first come first served basis. Since there are large numbers of potential participants, we will need to know by that early date (September 1) whether or not you plan on attending so we can make appropriate logistical arrangements. A $100 registration fee per attendee will be collected at the workshop. Attendees to the workshop will be largely self-funded similar to the IPCC model analysis workshop held in Hawaii in March, 2005.

We look forward to this opportunity to have WG1 LAs and CLAs from all four assessments gather as a group for a science meeting for the first time in the history of the IPCC. The outcomes from this unique workshop will provide important scientific direction as input to the early planning stages for the IPCC AR5.
Best regards from the organizing committee,

Gerald Meehl, Jonathan Overpeck, Susan Solomon, Thomas Stocker, and Ron Stouffer

# Appendix 2

Last year, a TV program opposing global warming alarmism, The Great Global Warming Swindle, was aired by channel 4 in Britain. The IPCC brought a complaint against the producers of the program to the British Office of Communications (OfCom). The OfCom held that the producers did not give the IPCC sufficient time to respond (they were given about a week), but that the program did not materially mislead the public. Steven McIntyre on his web site, www.climateaudit.org, analyzes the decision as well as the dishonest responses of the IPCC officials to the OfCom findings. It is a lovely example of self-refutation. That is to say, the IPCC officials demonstrated that they were acting in a political capacity in the very process of denying this.

## *Climate Science: Is it currently designed to answer questions?*

**Ofcom: The IPCC Complaint**

By Steve McIntyre

Ofcom's disposition of the IPCC Complaint is here page 43. There are many interesting aspects to this decision that are distinct from any of the others. Ofcom's actual finding is extremely narrow. IT rejected 2 of 6 complaints. On 3 of 6, it determined that the producers had provided notice to IPCC but the notice on Feb 27, 2007 did not leave IPCC with "reasonable time" to respond prior to the airing on March 8, 2007 (though Ofcom itself states that "three working days" is a "reasonable time" for the parties to file an appeal of the present decision. They also determined that the producers failed to give IPCC adequate notice that someone in the production would say that they were "politically driven". Had the producers sent their email of Feb 27, 2007 on (say) Feb 20, 2007, including a mention in the email that one of the contributors stated that IPCC was "politically driven", then the Swindle producers would appear to have been immune from the present findings. Little things do matter.

The two rejected claims are themselves rather interesting and make you scratch your head. As discussed below, Swindle contributors were said to have claimed that IPCC had predicted climate disaster and the northward migration of malaria as a result of global warming. IPCC denied ever making such claims and apparently felt that its reputation was sullied by being associated with such claims. These two matters were decided on other grounds, but many readers will be interested to read more about IPCC disassociating itself from claims that global warming would cause northward migration of malaria or predictions of climate disaster.

In addition, in its complaint, IPCC made grandiose claims about its "open and transparent process" and the role of review editors, describing the process as being in the public domain and by its nature designed to avoid "undue influence" of any reviewer. This will come as somewhat of a surprise to CA readers, who are familiar with the avoidance of IPCC procedures by Ammann and Briffa and the seemingly casual performance of review editor Mitchell and who have been following the relentless stonewalling by IPCC and IPCC officials of requests for specific information pertaining to this allegedly "open and transparent process".

Two Rejected Complaints
They discarded two parts of the complaint entirely.

IPCC denied that it had claimed that malaria "will" spread as a result of global warming (as stated by Channel 4) and said that it was unfair for Channel 4 to have broadcast this claim without their having an adequate opportunity to respond. The claim was decided on other grounds (that the allegation by Paul Reiter did not mention specifically mention IPCC). However, many readers will be surprised and interested to know that IPCC considers that its reputation is diminished by attributing to it the view that malaria will spread as a result of global warming.

## *Climate Science: Is it currently designed to answer questions?*

IPCC complained that the "programme falsely claimed that its FAR (1990) predicted "climatic disaster as a result of global warming" without an opportunity to defend itself against the indignity of being accused of making such a claim. It's a relief to the rest of us to know that not only is the IPCC not predicting climatic disaster, but it considers being associated with such a claim to be an insult. Ofcom considered some interesting contemporary evidence, including a speech by Margaret Thatcher, the scientific content of which was approved by Houghton, and came to the view that this was not an unreasonable characterization. Their decision on this issue stated:

> the Committee considered that the comment that described the FAR (1990) as predicting "climatic disaster as a result of global warming" was not an allegation against the IPCC and was not unfair to it. It was not, therefore, incumbent on the programme makers to have offered the IPCC an appropriate and timely opportunity to respond to this particular comment.

The most interesting part of these two issues were the IPCC defences.

**Three Issues where the notice was insufficiently timely**

On three parts of the Complaint (Reiter's criticism of the malaria section of the IPCC report, Reiter's criticism of how IPCC made up its author lists, Seitz' criticism of the SAR-Santer fiasco), Ofcom found that Swindle had provided notice to IPCC within the requirements, but had failed to provide IPCC with enough time to respond.

What would be a reasonable amount of time? Ofcom says in their Guidelines for the handling of standards complaints and cases (in programmes and sponsorship) that three working days is a "reasonable time" for an appeal, 5 working days for broadcasters to deliver any requested material and 10 working days to deliver certain sorts of detailed written submissions.

While the producers had preliminary contact with IPCC in October 2006 (as a result of which they were referred to a website), the first notice to IPCC that they would be presenting the Reiter and Seitz allegations came on Feb 26, 2007 (a Monday). to which there was no response. A follow-up email was sent three days later on March 1, 2007, again with no response. At the time of the show's first airing on March 8, 2007, ten days (8 working days) after the first notice letter, IPCC had still sent no response. Nor did it send one prior to the second airing. Ofcom noted:

> *the IPCC is a large organisation with considerable resources at its disposal and that it employs a dedicated Information and Communications Officer. On the face of it, these factors might be taken to suggest the IPCC should have been in a position to respond to the programme makers' emails (subject to being provided with sufficient information about the*

### Climate Science: Is it currently designed to answer questions?

*allegations that would be made in the programme)*

On the other hand, Ofcom noted that the producers had failed to properly inform IPCC of the deadlines:

> As mentioned above, it was significant that the programme maker's email of 26 February 2007 gave the IPCC no indication of when its response was required and the follow-up email of 1 March 2007 (sent at 7.33pm) subsequently gave a deadline of the following day. Neither of these emails indicated the date of broadcast.
>
> Taking into account all the above factors, the Committee considered that it was unreasonable for the programme makers to have expected the IPCC to understand that its response was required in a matter of days, and that it was not reasonable to expect the IPCC to be able to provide a response within the one day of being advised of the deadline. The Committee therefore found that the opportunity to respond had not been offered in a timely way.

On these particular findings, there's a process lesson about the need for clear and unequivocal notice. In this particular case, it seems highly unlikely that IPCC was going to bother responding in any event. So the producers could easily have avoided this particular problem merely by giving clearer and somewhat more informative notice. For example, had they sent out the email on Feb 20, 2007 instead of Feb 27, 2007, notifying the IPCC of their deadline, then it's hard to see how these parts of the IPCC complaint could have even got as far as they did.

I note that it appears that IPCC itself did not even file the "IPCC Complaint". It appears to be another concoction by Rado and associates. Their website says that:
>    *Sir John Houghton … co-authorised our Fairness complaint on behalf of the IPCC…. Dr Pachauri co-authorised our Fairness complaint on behalf of the IPCC. …Martin Parry also co-authorised our Fairness complaint on behalf of the IPCC… Professor [Robert] Watson co-authorised our Fairness complaint on behalf of the IPCC."*

which I take this as evidence that IPCC itself did not author the complaint. Normally, in order to be heard by Ofcom, a "fairness" complaint has to be made by the person directly affected. There are situations in which a third party can be authorized to make the complaint; I haven't examined whether these situations apply here.

However the form of IPCC "authorization" seems highly curious. John Houghton supposedly "co-authorised our Fairness complaint on behalf of the IPCC". While Houghton has obviously been an important figure in the IPCC movement, he is not listed at the IPCC website as one of its present officers and would not appear to have sufficient current authority to "authorize" the complaint. Robert Watson's appearance on this list is also interesting. Watson is likewise not listed as an current IPCC officer; Rado's website states that Watson is currently DEFRA's Chief Scientific Adviser. That a DEFRA employee should perceive himself as having the authority to authorize the commencement of an action in the

### *Climate Science: Is it currently designed to answer questions?*

U.K. on behalf of IPCC, which, under other circumstance, asserts its immunity rights as an international organization, is intriguing to say the least.

**A "Political" Organization**

The last "issue" in play was the statement by Philip Stott that IPCC was a "politically driven" organization.

Dr Philip Stott: "The IPCC, like any UN body, is political. The final conclusions are politically driven."

This matter differed somewhat from the 3 considered under the previous head in that no notice was given to the IPCC in their Feb 26, 2007 email that the production would say that they are "political".

In its defence, Channel 4 said

*The programme contributor, Dr Philip Stott, was merely making a statement of fact. Channel 4 said the programme made the important and valid point that the IPCC is political as well as scientific. Channel 4 said the IPCC chairmen and authors are nominated by governments and the reports are viewed by government officials prior to publication. Further, Channel 4 said the IPCC had been criticised on a number of occasions for being hampered by political interference. Channel 4 therefore maintained it was entirely fair for Professor Stott to state that the IPCC is "politically driven".*

The IPCC response will be particularly intriguing to Climate Audit readers who have followed IPCC's refusal to provide a complete archive of its Review Comments and Responses (in direct breach of their own formal procedures), a refusal abetted by corresponding refusals of national FOI requests. Ofcom summarizes their response:

*In relation to the IPCC being "politically driven", the IPCC said that the requirement for openness and transparency in its processes ensured that it was impossible for any undue interference to take place or any undue pressure to be applied by any reviewer (government or otherwise).*

*The IPCC said the government expert reviewer is free to ask any lead author to reconsider what they have written, but based solely on scientific content. The lead author will then consider the comment or request for change. If the lead author then wishes to make the change, he/she has to account for the decision to his/her review editor, who will make the final decision. Such changes must then be documented and the results made public.*

*The IPCC said that, given the IPCC's own procedures, Channel 4's arguments in relation to this head of complaint were either ill-informed or disingenuous.*

### *Climate Science: Is it currently designed to answer questions?*

Huh? This is not a true description of the process that I've experienced or that has been documented here. "Disingenuous" - they must be taking etiquette lessons from Michael Mann.

In terms of my own personal experience, we know that Ammann evaded the formal "open and transparent" process by sending review comments about our work outside the properly instituted process and that the parties have subsequently refused to produce the presumably adverse comments. Did these exchanges result in "undue interference" or "undue pressure" by a reviewer? The purpose of the "open and transparent" process is to do what IPCC represented to Ofcom that it did. Too bad that it's not a true description.

Similarly with the role of the Review Editors. IPCC testified to Ofcom that the "review editor" made the final decision. But Review Editor Mitchell has said that these decisions were up to Briffa and the chapter authors. Although IPCC says here that this process is "public", IPCC has refused to provide Mitchell's comments and Mitchell has concocted absurd and untrue reasons to avoid producing the comments (even claiming that he acted as an IPCC review editor in a "personal" capacity and that he has destroyed all his IPCC correspondence).

Here's how Ofcom decided this matter:

> In the Committee's opinion, viewers would have understood from the full section (quoted above) that the IPCC was not a purely scientific body and that its 'scientific' conclusions were significantly tainted by political interests.
>
> The Committee considered that such an impression went to the core of the IPCC's function and reputation: in this regard it noted that the IPCC was set up following international governmental accord with the aim of producing objective scientific assessments to inform policy and decision making worldwide. The Committee considered that "politically driven" was a strong and potentially damaging allegation which, within the context of this part of the programme, suggested direct political influence and was clearly intended to call into question the credibility of the IPCC….
>
> … In the circumstances, the Committee concluded that the IPCC was not afforded a timely or appropriate opportunity to respond to the significant allegation that the conclusions of the IPCC were "politically driven". This resulted in unfairness to the IPCC in the programme as broadcast.

**Summary**

So what exactly did IPCC win? Ofcom said that the producers should have given them more adequate notice time for Reiter's allegations about the review of the malaria section and the listing of authors and for Seitz' allegations about SAR and for the assertion that they would say that IPCC was "politically driven".

### *Climate Science: Is it currently designed to answer questions?*

Did Ofcom opine on whether IPCC was giving good or bad reports? Nope. It stuck to knitting and rendered carefully reasoned decisions on whether the producers gave adequate notice to someone being criticized, as required under the Broadcasting Code.

**"Vindication"**

Now look at the crowing about this decision by IPCC officials.

Pachauri:

> *We are pleased to note that Ofcom has vindicated the IPCC's claim against Channel Four in spirit and in substance, and upheld most of the formal complaints made by those who respect the IPCC process. It is heartening to see that the review process of the IPCC, and the credibility of the publications of the IPCC were upheld, as was the claim that Channel Four did not give the Panel adequate time to respond to most of their allegations. The IPCC is an organization that brings together the best experts from all over the world committed to working on an objective assessment of all aspects of climate change. The relevance and integrity of its work cannot be belittled by misleading or irresponsible reporting. We express our appreciation of the Fairness Committee at Ofcom, and are satisfied with their rulings on this matter.*

Some of this is simply untrue. Ofcom did not "uphold" the review process of the IPCC or the credibility of IPCC publications. Neither did it trash them. It simply did not consider them. Pachauri is totally misrepresenting the decision.
Houghton:

> *The ruling today from Ofcom regarding the Great Global Warming Swindle programme has exposed the misleading and false information regarding the Intergovernmental Panel on Climate Change (IPCC) that was contained in that programme and that has been widely disseminated by the climate denying community. The integrity of the IPCC's reports has therefore been confirmed as has their value as a source of accurate and reliable information about climate change.*

Again, all completely untrue. The Ofcom decision did "not expose the misleading and false information" regarding IPCC nor did it "confirm the integrity of the IPCC reports". Nor did it endorse the programme nor did it trash the integrity of the reports. It didn't make any decision on them one way or another. It simply said that the producers failed to give IPCC enough notice to respond.

Robert Watson

> *I am pleased that Ofcom recognized the serious inaccuracies in the Global Warming Swindle and has helped set the record straight.*

Again untrue. Ofcom did nothing of the sort. It made no attempt whatever to sort out the

*Climate Science: Is it currently designed to answer questions?*

scientific disputes.

Martin Parry:

> *This is excellent news. People and policymakers need to have confidence in the science of climate change. The reputation of the IPCC as the source of dependable and high quality information has been fully upheld by this Ofcom ruling. Channel 4's Great Global Warming Swindle was itself a disreputable attempt to swindle the public of the confidence it needs in scientific advice.*

Again completely untrue. The Ofcom ruling did not "uphold" the "reputation of the IPCC as the source of dependable and high quality information". Nor did it disparage its reputation. It simply said that IPCC didn't get enough time to respond.

# Appendix 3

From the *Boston Globe*

### Convincing the climate-change skeptics

### By John P. Holdren | August 4, 2008

THE FEW climate-change "skeptics" with any sort of scientific credentials continue to receive attention in the media out of all proportion to their numbers, their qualifications, or the merit of their arguments. And this muddying of the waters of public discourse is being magnified by the parroting of these arguments by a larger population of amateur skeptics with no scientific credentials at all. Long-time observers of public debates about environmental threats know that skeptics about such matters tend to move, over time, through three stages. First, they tell you you're wrong and they can prove it. (In this case, "Climate isn't changing in unusual ways or, if it is, human activities are not the cause.") Then they tell you you're right but it doesn't matter. ("OK, it's changing and humans are playing a role, but it won't do much harm.") Finally, they tell you it matters but it's too late to do anything about it. ("Yes, climate disruption is going to do some real damage, but it's too late, too difficult, or too costly to avoid that, so we'll just have to hunker down and suffer.")

All three positions are represented among the climate-change skeptics who infest talk shows, Internet blogs, letters to the editor, op-ed pieces, and cocktail-party conversations. The few with credentials in climate-change science have nearly all shifted in the past few years from the first category to the second, however, and jumps from the second to the third are becoming more frequent. All three factions are wrong, but the first is the worst. Their arguments, such as they are, suffer from two huge deficiencies.

*Climate Science: Is it currently designed to answer questions?*

First, they have not come up with any plausible alternative culprit for the disruption of global climate that is being observed, for example, a culprit other than the greenhouse-gas buildups in the atmosphere that have been measured and tied beyond doubt to human activities. (The argument that variations in the sun's output might be responsible fails a number of elementary scientific tests.)

Second, having not succeeded in finding an alternative, they haven't even tried to do what would be logically necessary if they had one, which is to explain how it can be that everything modern science tells us about the interactions of greenhouse gases with energy flow in the atmosphere is wrong.
Members of the public who are tempted to be swayed by the denier fringe should ask themselves how it is possible, if human-caused climate change is just a hoax, that: The leaderships of the national academies of sciences of the United States, United Kingdom, France, Italy, Germany, Japan, Russia, China, and India, among others, are on record saying that global climate change is real, caused mainly by humans, and reason for early, concerted action. This is also the overwhelming majority view among the faculty members of the earth sciences departments at every first-rank university in the world.

All three of holders of the one Nobel prize in science that has been awarded for studies of the atmosphere (the 1995 chemistry prize to Paul Crutzen, Sherwood Rowland, and Mario Molina, for figuring out what was happening to stratospheric ozone) are leaders in the climate-change scientific mainstream. US polls indicate that most of the amateur skeptics are Republicans. These Republican skeptics should wonder how presidential candidate John McCain could have been taken in. He has castigated the Bush administration for wasting eight years in inaction on climate change, and the policies he says he would implement as president include early and deep cuts in US greenhouse-gas emissions. (Senator Barack Obama's position is similar.)

The extent of unfounded skepticism about the disruption of global climate by human-produced greenhouse gases is not just regrettable, it is dangerous. It has delayed - and continues to delay - the development of the political consensus that will be needed if society is to embrace remedies commensurate with the challenge. The science of climate change is telling us that we need to get going. Those who still think this is all a mistake or a hoax need to think again.

*John P. Holdren is a professor in the Kennedy School of Government and the Department of Earth and Planetary Sciences at Harvard and the director of the Woods Hole Research Center.*

*Climate Science: Is it currently designed to answer questions?*

*Climate Science: Is it currently designed to answer questions?*

*Climate Science: Is it currently designed to answer questions?*

***Climate Science: Is it currently designed to answer questions?***

*Climate Science: Is it currently designed to answer questions?*